# Reducing Road Vehicle Fuel Consumption by Exploiting Connectivity and Automation: A Literature Survey

Miguel Alvarez, Chu Xu, Manuel A. Rodriguez, Abdullah Al-Mamun, Mohamed Wahba,
Sean Brennan, and Hosam K. Fathy
Department of Mechanical and Nuclear Engineering
The Pennsylvania State University
E-mail: hkf2@psu.edu

This paper examines the degree to which connectivity and automation can potentially reduce the overall fuel consumption of on-road vehicles. The paper begins with a simulation study highlighting the tradeoff between: (i) the fuel that a vehicle can save through speed trajectory shaping, versus (ii) the additional inter-vehicle spacing needed for this trajectory shaping to be feasible. This study shows that connectivity and automation are essential, rather than merely useful, for substantial reductions in the fuel consumed by fixed on-road vehicle powertrain/chassis configurations in traffic. Motivated by this insight, we survey the literature on the fuel savings achievable through different connected/automated vehicle technologies. This includes optimal vehicle routing, eco-arrival/departure at intersections, platooning, speed trajectory optimization, predictive driveline disengagement, predictive gear shifting, and predictive powertrain accessory control. This survey shows that the ability to shape vehicle speed trajectories collaboratively plays a dominant role in reducing urban/suburban fuel consumption, while platooning plays a dominant role in influencing the attainable fuel savings on the highway. Moreover, the survey shows that the degree to which connectivity/automation can reduce on-road vehicle fuel consumption, in both urban/suburban and highway settings, depends critically on the integration of powertrain- and chassis-level control.

Topics: Powertrain Control, Energy Management, Vehicle Connectivity/Automation

## 1. INTRODUCTION

This paper examines the degree to which connectivity, automation, and autonomy can help road vehicles reduce their fuel consumption. The paper is motivated by the growing interest in connectivity and automation in today's automotive market. The driving forces behind this interest are diverse, and include: (i) improving vehicle safety, (ii) easing traffic congestion, and (iii) reducing vehicle fuel consumption. These driving forces are mutually coupled, in the sense that there are non-trivial synergies and tradeoffs between them. For example, the use of vehicle connectivity to ease traffic congestion can have significant fuel consumption implications.

The fuel consumption of any road vehicle depends on three factors, namely: (i) its powertrain dynamics, (ii) its chassis dynamics, and (iii) the duty cycles imposed on its powertrain/chassis. Each of these factors suggests a path towards fuel consumption reduction. Powertrain and chassis design innovations such as hybridization and "light-weighting", for example, have significant roles to play in reducing vehicle fuel consumption. Our focus in this paper is on the degree to which connectivity and automation, coupled with predictive control, can reduce the fuel consumption of a "fixed" or "given" road vehicle. This is an inherently conservative focus, in the sense that it neglects the additional fuel economy benefits of exploiting connectivity/automation to design more fuel-efficient vehicle chassis and powertrains.

Our survey of the vehicle connectivity/automation literature is guided by insights from a simple simulation study. The study highlights the sheer magnitude of the additional inter-vehicle spacing that may be needed when vehicles shape their individual speed trajectories for fuel consumption without inter-vehicle communication/ coordination. A key conclusion of this study is that connectivity and automation are essential, rather than merely useful, if one seeks to achieve the best possible fuel consumption levels for a fixed/given powertrain/chassis design. Motivated by this insight, we survey the literature on the fuel savings achievable through different connected/automated vehicle (CAV) technologies. This survey covers technologies such as optimal vehicle routing, eco-arrival/departure at intersections, platooning, speed trajectory optimization, predictive driveline disengagement, predictive gear shifting, and predictive powertrain accessory control. The paper concludes with a summary of key insights from this survey, paying particular attention to the degree to which different CAV technologies are valuable in urban/suburban versus highway driving.

## 2. MOTIVATING STUDY

Consider the problem of "smoothing" the drive cycle of a conventional, automatic transmission vehicle to improve its fuel economy. The particular vehicle examined here is a medium-duty truck modeled in earlier work by the authors [1], but the insights from this study are more broadly applicable. The truck's longitudinal speed, $v(t)$, is governed by a point-mass chassis model:

$$\dot{v} = \frac{1}{M}\left\{F_{prop} - F_{brake} - \frac{1}{2}\rho C_d A_f v^2 - \mu M g\right\}, \quad (1)$$

where $M$ is the truck's mass, $F_{prop}$ is the effective propulsion or traction force, $F_{brake}$ is the effective braking force, $\rho$ is the density of air, $C_d$ is the truck's drag coefficient, $A_f$ is the truck's frontal area, $\mu$ is a rolling resistance coefficient, and $g$ denotes gravity. The truck's automatic transmission selects its gear ratio, $R$, based on driver pedal command and vehicle speed using a simple shift map. This leads to the following equations:

$$F_{prop} = \frac{1}{R_w}\eta R \tau_t, \quad \omega_t = \frac{1}{R_w} R v, \quad (2)$$

where $R_w$ is the truck's tire radius, $\eta$ is its effective transmission efficiency, $\tau_t$ is the torque generated by its torque converter's turbine, and $\omega_t$ is the turbine's angular velocity. Denoting the propulsion torque and speed provided at the truck's engine shaft by $\tau_e$ and $\omega_e$, respectively, one obtains:

$$\dot{\omega}_e = \frac{1}{J_e}(\tau_e - \tau_i), \quad \tau_{i,t} = f_{1,2}(\omega_i, \omega_t), \quad (3)$$

where $J_e$ is an effective engine output shaft inertia (including flywheel inertia), and the functions $f_{1,2}$ capture the static characteristics of the truck's torque converter. This includes both the torque ratio vs. speed ratio relationship and the capacity factor vs. speed ratio relationship for the torque converter. Finally, the rate of engine fuel consumption, $\dot{m}_f$, is governed by:

$$\dot{m}_f = (\tau_e \omega_e + P_{i/a})\beta(\tau_e + P_{i/a}/\omega_e, \omega_e), \quad (4)$$

where $P_{i/a}$ is the power demand placed on the engine by the idling system and other accessories, and $\beta$ is the engine's brake-specific fuel consumption as a function of total engine torque and engine speed.

The above truck model receives two inputs from the corresponding driver model, namely: an engine torque command plus vehicle braking. The driver model uses PI control to match a desired speed trajectory as a function of time. Examples of such a trajectory include the well-known HWFET and LA92 drive cycles.

Fuel consumption, particularly for conventional powertrains, depends strongly on the smoothness of the underlying drive cycles. Drive cycles with more aggressive or frequent acceleration/braking events, including stop-and-go events, inherently require more fuel. With this in mind, consider the problem of "smoothing" a given drive cycle for the above truck. One simple approach for such smoothing is to use a moving average filter. Suppose, for instance, that the original drive cycle dictates a velocity setpoint $v_k^*$ at every sampling instant, $k$. Then one can use a moving average filter to create a new velocity setpoint, $v_k'$, as follows:

$$v_k' = \frac{1}{2m+1}\sum_{i=k-m}^{k+m} v_i^*, \quad (5)$$

where $2m + 1$ represents the total number of sampling times used for constructing the moving average. The tuning of this filter's parameters has a direct impact on the resulting vehicle fuel consumption. Intuitively, at least for a vehicle with a conventional powertrain, one may expect vehicle fuel economy to improve as the filter generates smoother cycles, with more benign start-stop behavior. This can be achieved by increasing the value of $m$, but it comes at a price, namely, a growing deviation of the vehicle's speed trajectory from the original standard drive cycle. In a hypothetical scenario, where other vehicles on the road do adhere to the original standard cycle, this deviation translates to an additional amount of inter-vehicle spacing needed in order for the above cycle smoothing to be feasible. Thus, in the absence of vehicle connectivity/automation, there is a fundamental tradeoff between the fuel economy improvements associated with drive cycle smoothing versus the additional inter-vehicle spacing required for such smoothing.

Figures 1 and 2 illustrate the above tradeoff, for both the HWFET and LA92 drive cycles, respectively. A few insights are evident from these figures. First, speed trajectory optimization, alone, can reduce the fuel consumption of an internal combustion vehicle significantly. This is particularly true in aggressive stop-and-go traffic, where cycle smoothing alone can reduce fuel consumption by almost 30% (as opposed to 4.5% on the highway). Second, the above benefit increases considerably with the ability to communicate and coordinate not just with other vehicles, but also with traffic signals. This is evident from the contrast between the "stars" tradeoff front (where the vehicle can smooth its speed trajectory at every time instant) and the "circle" tradeoff front (where smoothing is only allowed when the original drive cycle indicates a non-zero target speed). Third, without inter-vehicle coordination, the additional inter-vehicle spacing needed for the above fuel consumption reduction levels is infeasible, particularly in congested traffic. Altogether, these insights lead to the following conclusion: connectivity and automation are critical prerequisites, rather than mere enablers, for substantial reductions in fuel consumption for fixed on-road vehicle powertrain/chassis configurations in traffic.

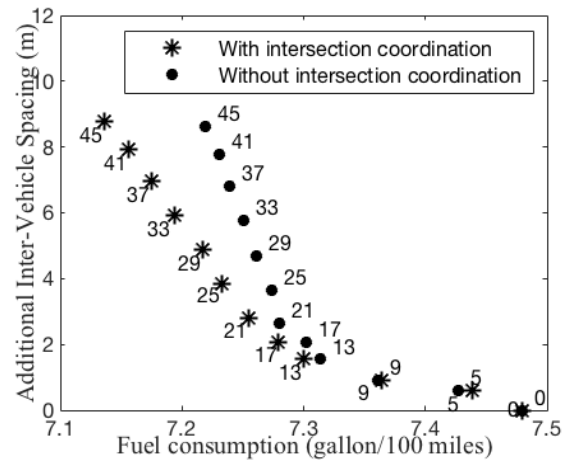

Fig. 1: Tradeoff between fuel savings and additional inter-vehicle spacing needs, both with (stars) and without (circles) intersection coordination, HWFET

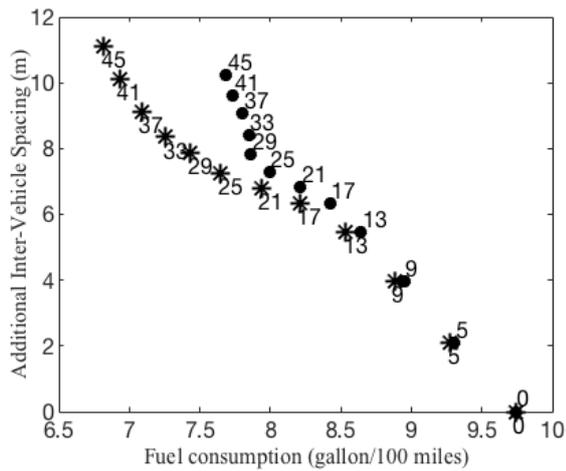

Fig. 2: Tradeoff for LA92 Cycle

## 3. LITERATURE SURVEY

The literature on fuel consumption minimization in connected and automated vehicles is very extensive. One method for navigating this literature is to examine the main CAV technologies discussed within it, including:

### 3.1 Optimal Vehicle Routing

Choosing a road vehicle's route can affect its trip distance, trip time, and trip fuel consumption significantly [2,3]. Recognizing this, the literature generally tackles optimal routing as a multi-objective problem [3]. Optimal routing is inherently a stochastic problem, owing to the uncertainties associated with factors such as traffic patterns. The literature addresses this stochasticity using strategies such as allowing the edge costs in a routing problem to be time-dependent and/or uncertain [4]. This increases the computational cost of the optimal routing problem somewhat. However, the literature is still successful in solving routing problems in a computationally tractable manner, and also in integrating it with both hybrid vehicle power management and platooning [2,5]. Routing is a promising CAV technology, with significant potential fuel consumption reductions reporting in the literature. For instance, an extensive simulation case study reports a 5% reduction in fuel consumption due to coordinated platoon routing and planning [5]. It is important to recognize the potential fuel economy benefits of solving the optimal routing problem in both space and time whenever possible. For instance, future autonomous commercial freight vehicles may be able to reduce their fuel consumption significantly by shifting their trips to less-congested times of the day, as part of an optimal space/time routing problem.

### 3.2 Eco-Departure/Arrival at Intersections

Optimizing the interactions between vehicles and signalized intersections can reduce both congestion and fuel consumption substantially. This is particularly true for conventional (as opposed to hybrid) vehicle powertrains, for which "stop-and-go" driving repeatedly consumes fuel to build up kinetic energy, then dissipates this energy irreversibly. The exchange of information, both among vehicles and between vehicles and traffic signals, makes it possible to predict/avoid events such as abrupt stops at red lights [6]. This can reduce overall vehicle fuel consumption significantly, particularly for conventional vehicles in urban/suburban settings. Fuel savings well in excess of 40% are reported in the literature for the eco-departure and eco-arrival of vehicles at intersections [7]. While the precise fuel savings from this CAV technology depend on the underlying traffic patterns, the savings remain quite substantial for many different traffic patterns. Moreover, these substantial fuel savings are corroborated by both simulation-based and experimental studies, with one study showing 14% fuel savings for a passenger car in both types of studies [8]. Motivated by these significant potential fuel savings, researchers explore multiple approaches for eco-departure/arrival at intersections, grounded in ideas from the reservation-based scheduling [9], cooperative control [10,11], and model predictive control literatures [12,13]. This includes the use of tools such as pseudo-spectral optimization to ensure computational tractability [14]. It also includes the use of tools such as evolutionary and/or hierarchical optimization to assist vehicles in forming, joining, and re-organizing platoons in the vicinity of signalized and/or intelligent intersections [15,16].

### 3.3 Platooning

Traveling in a platoon allows a vehicle to lower its aerodynamic drag coefficient and reduce its fuel consumption. Experiments estimate a fuel saving of 4-7% for platoons of heavy duty vehicles [17,18]. There is, however, a fundamental tradeoff between the aerodynamic drag reduction and the increase in braking to maintain a platoon, particularly when terrain changes. One way the literature addresses this issue is to use terrain information in a receding horizon framework to optimize the trajectories of the vehicles [19-21]. This can be done cooperatively [19,20], where the trajectories of the vehicles are shaped to reduce the fuel consumption of the platoon. It can also be done non-cooperatively [21], where the follower vehicles use knowledge of the future trajectory to minimize unnecessary braking and reduce fuel consumption. Another part of the literature focuses on the formation of platoons [5,22-26]. More specifically, the above studies focus on the coordination of different vehicles' trips to ensure the formation of platoons along their route. The literature is divided between centralized coordinators [22,23,5,24], and de-centralized coordinators [25,26] that use game-theoretic approaches to the formation of platoons.

### 3.3 Speed Trajectory Optimization

Knowing the road elevation and the traffic congestion patterns of a road ahead can help shape the speed trajectory of a vehicle to minimize fuel consumption. Using terrain information in a model predictive control framework [27-30] can yield fuel savings of 3.5% for a heavy-duty vehicle on a given road. Analytical approaches based on Pontryagin's maximum principle yield fuel savings of 5-10% [31,32]. The literature also exploits knowledge of the traffic ahead to provide the speed trajectory that minimizes fuel [33-37]. There are different approaches to estimating the traffic ahead. One way is to deterministically model the traffic ahead and implement it in a model predictive control framework [33,34,35]. Experiments show that this approach can yield 15.8% savings in fuel consumption for a passenger car [37]. Another approach is to use a probability distribution to model the headway constraint relative to the leading vehicle [36].

### 3.4 Optimal Powertrain Control

Vehicle connectivity and automation technologies create important opportunities for saving fuel at the powertrain level. These opportunities can be seen from the contrast between connected/automated vehicles on the one hand, and more traditional human-driven road vehicles on the other hand. Predicting the propulsion power demanded by a human driver can be quite challenging. As a result, traditional vehicle powertrain control algorithms must achieve a certain degree of robustness to uncertainties in human driver demands. Connected and automated vehicles, in contrast, plan and optimize their future routes and speed trajectories. Information from this chassis-level optimization is then available for predictive powertrain control. Fuel savings opportunities exist both when chassis-level optimization feeds into powertrain-level optimization and when the two optimization problems are solved simultaneously, with the latter *co-optimization* approach leading to greater potential fuel savings.

There are at least three key opportunities for saving fuel through optimal powertrain control in connected/automated vehicles, namely: (i) predictive driveline disengagement, (ii) predictive gear shifting, and (iii) predictive accessory control. Much of the driveline disengagement literature builds on the use of "pulse and glide" (PnG) driving for adaptive cruise control on roads with known upcoming terrains [38,39-41]. PnG driving is a dynamic vehicle speed trajectory planning/control policy. It involves switching periodically between acceleration events at low engine brake-specific fuel consumption (bsfc) levels and coasting events where the vehicle driveline is potentially disengaged. PnG driving can achieve significant fuel saving in both continuously-variable [40] and step-gear mechanical transmission [41] by switching the engine periodically between the minimum BSFC point and the idling point.

There is a rich existing literature focusing on the optimal predictive scheduling of vehicle gear shifting for objectives such as fuel consumption minimization. The work in this literature uses a variety of optimization algorithms, including dynamic programming [42,43,44,45], Pontryagin methods [45], model predictive control [46], and fuzzy logic [47]. Studies consistently show that optimal gear shifting can reduce fuel consumption significantly, both in simulation and experimentally. Predictive gear shift scheduling using rule-based control schemes extracted from dynamic programming studies can reduce vehicle fuel consumption by 1% on the highway, and 1-2% in urban settings [44]. Moreover, heavy-duty trucks experiments show fuel consumption reductions of 1.91%-6.2% through a combination of optimal predictive gear scheduling and cruise control [28].

The third key powertrain-level strategy for fuel consumption reduction in connected/automated vehicles is predictive accessory control. The idea, here, is to schedule the usage of vehicle accessories during braking events, thereby using the vehicle's kinetic energy to drive the accessories instead of dissipating this energy. The optimal control of accessories like air conditioning systems can reduce fuel consumption by 1-2% over a broad range of driving scenarios [48].

### 4. DISCUSSION AND CONCLUSIONS

Figure 3 summarizes this literature survey by presenting the degree to which different technologies have the potential to contribute to overall fuel savings in urban/suburban and highway driving. A strong contrast is visible between urban/suburban and highway driving scenarios. On the one hand, platooning is the single most beneficial vehicle connectivity/autonomation technology for highway fuel savings, particularly for medium- and heavy-duty trucks. On the other hand, the benefits of platooning are negligible (and not shown in the figure) for urban/suburban settings. In contrast, the ability to collaborative shape vehicle duty cycles through speed trajectory optimization and the optimal departure/arrival at intersections dominates the potential fuel savings on those settings. On the whole, the potential for fuel consumption reduction in connected/automated vehicles is quite promising. However, one important caveat is that the degree to which future vehicles will realize this potential depends, critically, on how successful they are in integrating chassis- and powertrain-level optimal control.

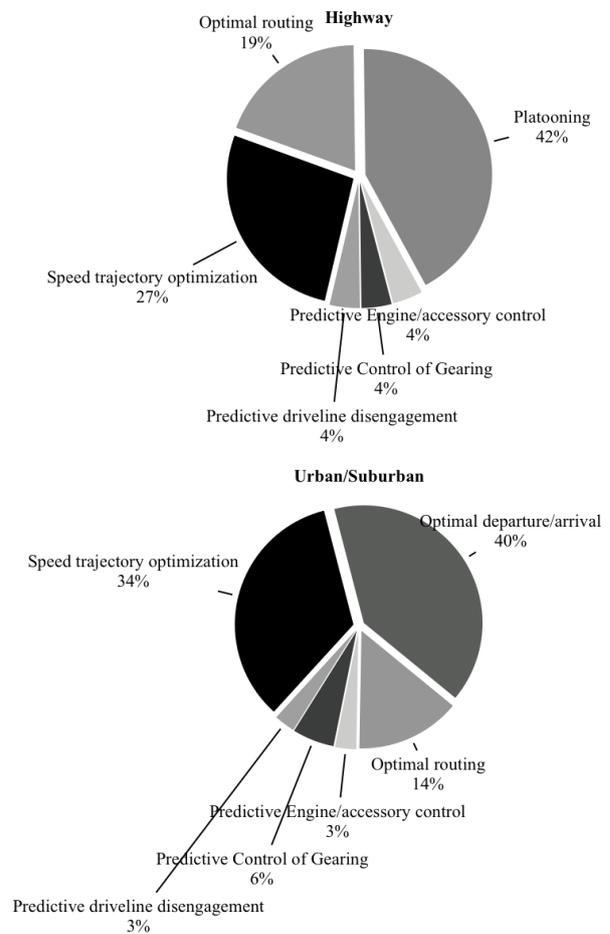

Fig. 3: Impact of different connectivity/automation technologies on potential fuel savings

**ACKNOWLEDGEMENTS**

The authors gratefully acknowledge this work's support/funding by the ARPA-E NEXTCAR program.

**REFERENCES**

[1] Fathy, H., Ahlawat, R., and Stein, J., "Proper Powertrain Modeling for Engine-in-the-Loop Simulation", Proc. of IMECE, 2005, pp. 1195-1201.

[2] Guo, C., Yang, B., Andersen, O., Jensen, C. S., and Torp, K., "EcoSky: Reducing Vehicular Environment Impact through Eco-Routing", Proc. of Int. Conf. on Data Eng., 2015, pp. 1412-1415.

[3] Yang, B., Guo, C., Jensen, C. S., Kaul, M., and Shang, S., "Stochastic Skyline Route Planning Under Time-Varying Uncertainty", Proc. of Int. Conf. on Data Eng., 2014, pp. 136-147.

[4] Jurik, T., Cela, A., Hamouche, R., Natowicz, R., Reama, A., Niculescu, S.I. and Julien, J., "Energy Optimal Real-Time Navigation System", IEEE Intel. Trans. Sys. Mag., vol. 6, no. 3, 2014, pp.66-79.

[5] Besselink, B., Turri, V., van de Hoef, S.H., Liang, K.Y., Alam, A., Mårtensson, J. and Johansson, K.H., "Cyber–Physical Control of Road Freight Transport", Proc. of the IEEE, vol. 104, no. 5, 2016, pp.1128-1141.

[6] Mandava, S., Boriboonsomsin, K., and Barth, M., "Arterial Velocity Planning Based on Traffic Signal Information Under Light Traffic Conditions", Proc. of IEEE Conf. on Int. Trans. Sys., 2009, pp. 160-165.

[7] Asadi, B., and Vahidi, A., "Predictive Cruise Control: Utilizing Upcoming Traffic Signal Information for Improving Fuel Economy and Trip Time", IEEE Trans. on Control Sys. Tech., vol. 19, no. 3, 2011, pp. 707-714.

[8] Xia, H., Boriboonsomsin, K., Schweizer, F., Winckler, A., Zhou, K., Zhang, W.B. and Barth, M., "Field Operational Testing of Eco-Approach Technology at a Fixed-Time Signalized Intersection", Prof. Int. Conf. on Intel. Trans. Syst., 2012, pp. 188-193.

[9] Huang, S., Sadek, A. W., and Zhao, Y., "Assessing the Mobility and Environmental Benefits of Reservation-Based Intelligent Intersections Using an Integrated Simulator," IEEE Trans. Intel. Transp. Syst., vol. 13, no. 3, 2012, pp. 1201–1214.

[10] Zohdy, I. H., and Kamalanathsharma, R. K., "Intersection Management for Autonomous Vehicles using iCACC," Proc. IEEE International Conf. on Intel. Trans. Sys., 2012, pp. 1109–1114.

[11] Lee, J., Park, B., Malakorn, K., and So, J., "Sustainability Assessments of Cooperative Vehicle Intersection Control at an Urban Corridor," Transp. Res. Part C Emerg. Technol., vol. 32, no. SI, 2013, pp. 193–206.

[12] Kamal, M. A. S., Mukai, M., Murata, J., and Kawabe, T., "Model Predictive Control of Vehicles on Urban Roads for Improved Fuel Economy," IEEE Trans. Control Syst. Technol., vol. 21, no. 3, 2013, pp. 831–841.

[13] HomChaudhuri, B., Vahidi, A., and Pisu, P., "A Fuel Economic Model Predictive Control Strategy for a Group of Connected Vehicles in Urban Roads," Proc. Am. Control Conf., 2015, pp. 2741–2746.

[14] He, X., Liu, H. X., and Liu, X., "Optimal Vehicle Speed Trajectory on a Signalized Arterial with Consideration of Queue," Transp. Res. Part C Emerg. Technol., vol. 61, 2015, pp. 106–120.

[15] Tallapragada, P. and Cortés, J., "Hierarchical-Distributed Optimized Coordination of Intersection Traffic," arXiv preprint, 2016, pp. 1–13.

[16] Liu, B., and El Kamel, A., "V2X-Based Decentralized Cooperative Adaptive Cruise Control in the Vicinity of Intersections," IEEE Trans. Intell. Transp. Syst., vol. 17, no. 3, 2016, pp. 644–658.

[17] A. A. Alam, A. Gattami, and K. H. Johansson, "An experimental study on the fuel reduction potential of heavy duty vehicle platooning," Intell. Transp. Syst. (ITSC), 2010 13th Int. IEEE Conf., pp. 306–311, 2010.

[18] A. Alam, J. Mårtensson, and K. H. Johansson, "Experimental evaluation of decentralized cooperative cruise control for heavy-duty vehicle platooning," Control Eng. Pract., vol. 38, pp. 11–25, 2015.

[19] A. Alam, J. Mårtensson, and K. H. Johansson, "Look-ahead cruise control for heavy duty vehicle platooning," IEEE Conf. Intell. Transp. Syst. Proceedings, ITSC, no. Itsc, pp. 928–935, 2013.

[20] V. Turri, B. Besselink, and K. H. Johansson, "Cooperative look-ahead control for fuel-efficient and safe heavy-duty vehicle platooning," arXiv:1505.00447 [cs.SY], vol. Submitted, no. 1, pp. 1–16, 2015.

[21] V. Turri, Y. Kim, J. Guanetti, K. H. Johansson, and F. Borrelli, "A model predictive controller for non-cooperative eco-platooning," Proc. Am. Control Conf., pp. 2309–2314, 2017.

[22] K. Y. Liang, J. Mårtensson, and K. H. Johansson, "When is it fuel efficient for a heavy duty vehicle to catch up with a platoon?" IFAC Proc. Vol., vol. 7, no. PART 1, pp. 738–743, 2013.

[23] K.-Y. Liang, "Coordination and Routing for Fuel-Efficient Heavy-Duty Vehicle Platoon Formation," KTH School, 2014.

[24] S. Van De Hoef, K. H. Johansson, and D. V. Dimarogonas, "Coordinating Truck Platooning by Clustering Pairwise Fuel-Optimal Plans," IEEE Conf. Intell. Transp. Syst. Proceedings, ITSC, vol. 2015–Octob, pp. 408–415, 2015.

[25] M. Saeednia and M. Menendez, "A consensus-based algorithm for truck platooning," IEEE Trans. Intell. Transp. Syst., vol. 18, no. 2, pp. 404–415, 2017.

[26] F. Farokhi and K. H. Johansson, "A game-theoretic framework for studying truck platooning incentives," IEEE Conf. Intell. Transp. Syst. Proceedings, ITSC, no. Itsc, pp. 1253–1260, 2013.

[27] E. Hellström, A. Fröberg, and L. Nielsen, "A real-time fuel-optimal cruise controller for heavy trucks

AVEC'18

using road topography information," in SAE World Congress, 2006, vol. 2006, no. 724, pp. 1–10.

[28] E. Hellström, M. Ivarsson, J. Aslund, and L. Nielsen, "Look-ahead control for heavy trucks to minimize trip time and fuel consumption," IFAC Proc. Vol., vol. 5, no. PART 1, pp. 439–446, Feb. 2007.

[29] W. Huang, D. M. Bevly, S. Schnick, and X. Li, "Using 3D road geometry to optimize heavy truck fuel efficiency," IEEE Conf. Intell. Transp. Syst. Proceedings, ITSC, pp. 334–339, 2008.

[30] M. A. S. Kamal, M. Mukai, J. Murata, and T. Kawabe, "Ecological vehicle control on roads with up-down slopes," IEEE Trans. Intell. Transp. Syst., vol. 12, no. 3, pp. 783–794, 2011.

[31] B. Saerens, H. A. Rakha, M. Diehl, and E. Van den Bulck, "A methodology for assessing eco-cruise control for passenger vehicles," Transp. Res. Part D Transp. Environ., vol. 19, pp. 20–27, 2013.

[32] E. Ozatay et al., "Cloud-based velocity profile optimization for everyday driving: A dynamic-programming-based solution," IEEE Trans. Intell. Transp. Syst., vol. 15, no. 6, pp. 2491–2505, 2014.

[33] N. J. Kohut, J. K. Hedrick, and F. Borrelli, "Integrating traffic data and model predictive control to improve fuel economy," IFAC Proc. Vol., vol. 42, no. 15, pp. 155–160, 2009.

[34] M. A. S. Kamal, M. Mukai, J. Murata, and T. Kawabe, "On board eco-driving system for varying road-traffic environments using model predictive control," Proc. IEEE Int. Conf. Control Appl., pp. 1636–1641, 2010.

[35] A. Asadi, B., Zhang, C., and Vahidi, "The role of traffic flow preview for planning fuel optimal vehicle velocity," ASME Dyn. Syst. Control Conf., pp. 1–7, 2010.

[36] C. Zhang and A. Vahidi, "Predictive Cruise Control with Probabilistic Constraints for Eco Driving," Proc. ASME 2011 Dyn. Syst. Control Conf., pp. 1–6, 2011.

[37] R. Schmied, H. Waschl, and L. Del Re, "Extension and experimental validation of fuel efficient predictive adaptive cruise control," Proc. Am. Control Conf., vol. 2015–July, pp. 4753–4758, 2015.

[38] H. Koch-Groeber and J. Wang, "Criteria for Coasting on Highways for Passenger Cars," in SAE Technical Paper 2014-01-1157, 2014.

[39] C. Rostiti, S. Stockar, and M. Canova, "A Rule-Based Control for Fuel-Efficient Automotive Air Conditioning Systems," SAE Int., 2015.

[40] S. E. Li, K. Li, C. Ahn, and X. Hu, "Mechanism of vehicular periodic operation for optimal fuel economy in free-driving scenarios," IET Intell. Transp. Syst., vol. 9, no. 3, pp. 306–313, 2015.

[41] Xu, S. et al. (2015), "Fuel-optimal cruising strategy for road vehicles with step-gear mechanical transmission," IEEE Trans. Intell. Transp. Syst., 16(6), pp. 3496-3507.

[42] D. V. Ngo, T. Hofman, M. Steinbuch, A. Serrarens, and L. Merkx, "Improvement of fuel economy in power-shift automated manual transmission through shift strategy optimization - An experimental study," in 2010 IEEE Vehicle Power and Propulsion Conference, VPPC 2010, 2010, pp. 1–5.

[43] B. Škugor, J. Deur, and V. Ivanović, "Dynamic Programming-Based Design of Shift Scheduling Map Taking into Account Clutch Energy Losses During Shift Transients," 2016.

[44] D. Kim, H. Peng, S. Bai, and J. M. Maguire, "Control of integrated powertrain with electronic throttle and automatic transmission," IEEE Trans. Control Syst. Technol., vol. 15, no. 3, pp. 474–482, 2007.

[45] A. Fröberg and L. Nielsen, "Optimal fuel and gear ratio control for heavy trucks with piece wise affine engine characteristics," in IFAC Proceedings Volumes (IFAC-PapersOnline), 2007, vol. 5, no. PART 1, pp. 335–342.

[46] J. Bishop, A. Nedungadi, G. Ostrowski, B. Surampudi, P. Armiroli, and E. Taspinar, "An Engine Start/Stop System for Improved Fuel Economy," in SAE Technical Paper, 2007, vol. 2007-01–17, no. 724.

[47] Yamaguchi, H., Narita, Y., Takahashi, H. and Katou, Y., 1993. Automatic transmission shift schedule control using fuzzy logic(No. 930674). SAE Technical Paper.

[48] Zhang, Q., Stockar, S., & Canova, M. (2016). Energy-Optimal Control of an Automotive Air Conditioning System for Ancillary Load Reduction. IEEE Transactions on Control Systems Technology, 24(1), 67-80.